\documentstyle[aps,epsf,floats,amsfonts,amssymb,amsmath]{revtex}
\newcommand{\bea}{\begin{eqnarray}}\newcommand{\eea}{\end{eqnarray}}
\newcommand{\brr}{\begin{array}}\newcommand{\err}{\end{array}}
\newcommand{\bit}{\begin{itemize}}\newcommand{\eit}{\end{itemize}}
\newcommand{\ben}{\begin{enumerate}}\newcommand{\een}{\end{enumerate}}

\def\lan{\langle}
\def\lf{\left}

\def\non{\nonumber}
\def\ran{\rangle}

\def\ri{\right}

\def\al{\alpha}\def\bt{\beta}

\def\te{\theta}

\def\1{{_{1}}}
\def\2{{_{2}}}
\begin{document}

\title{Neutrino mixing and cosmological constant}
\author{M.~Blasone${}^{\flat\sharp}$, A.~Capolupo${}^{\flat}$,
S.~Capozziello${}^{\flat}$, S.~Carloni${}^{\flat}$ and
G.~Vitiello${}^{\flat \sharp }$ \vspace{3mm}}

\address{ ${}^{\flat}$ Dipartimento di  Fisica "E.R. Caianiello" and INFN,
Universit\`a di Salerno, I-84100 Salerno, Italy
\\ [2mm] ${}^{\sharp}$ Unit\`a INFM, Salerno, Italy
\vspace{2mm}}


\maketitle

\begin{abstract}

We report on the recent result
  that the non--perturbative vacuum structure associated
with neutrino mixing leads to a non--zero contribution to the
value of the cosmological constant. Its value is estimated by
using the natural cut--off appearing in the quantum field theory
formalism for neutrino mixing.

\end{abstract}

\vspace{8mm}

In this report we show  that the vacuum energy induced by  the
neutrino mixing may contribute to the value of cosmological
constant \cite{Blasone:2004yh}.

It is known that vacuum for neutrinos with definite mass is not
invariant under the field mixing transformation and in the
infinite volume limit it is unitarily inequivalent to the vacuum
for the neutrinos with definite flavor \cite{BV95}-\cite{aspects}.
This phenomenon is crucial in order to obtain a non--zero
contribution to the cosmological constant \cite{Blasone:2004yh};
it also
 affects the oscillation formula which turns out to be
different from the usual Pontecorvo formula \cite{Pontec}. In the
following, for simplicity we restrict ourselves to the two flavor
mixing and we use Dirac neutrino fields.


It was shown \cite{remarks} that in Quantum Field Theory (QFT) it
is possible to construct flavor states for neutrino fields. In the
limit of infinite volume, these states are orthogonal to the mass
eigenstates: we have two inequivalent vacua each other related by
the mixing generator $G_{\te}(t)$:
$|0(t)\ran_{f}\,\equiv\,G_{\te}^{-1}(t)\;|0\ran_{m}.$ Here, $\te$
is the mixing angle, $t$ is the time, $|0(t)\ran_{f}\ $ and
$|0\ran_{m},$ are the flavor and the mass vacua, respectively.
$G_{\te}(t)$ is given by: \bea G_{\bf \te}(t)=\exp\Big[\te\int
d^{3}{\bf
x}\lf(\nu_{1}^{\dag}(x)\nu_{2}(x)-\nu_{2}^{\dag}(x)\nu_{1}(x)\ri)
\Big]  . \eea

A Bogoliubov transformation is involved in connecting the flavor
annihilation operators to the mass annihilation operators. We
consider in particular the Bogoliubov coefficient $V_{\bf k}$
which is related to the condensate content of the flavor vacuum
\cite{BV95}: \bea\label{con} _{f}\langle 0| \al_{{\bf k},j}^{r
\dag} \al^r_{{\bf k},j} |0\rangle_{f}\,= \;_{f}\langle 0|
\bt_{{\bf k},j}^{r \dag} \bt^r_{{\bf k},j} |0\rangle_{f}\,=\;
|V_{{\bf k}}|^{2}\, \sin^{2}\te, \qquad \quad j=1,2,\eea where
$\al^r_{{\bf k},j}$, $\bt^r_{{\bf k},j}$ are the annihilation
operators for the neutrino fields $\nu_{1}$, $\nu_{2}$ with
definite masses, $m_{1}$, $m_{2}$ . $|V_{{\bf k}}|^{2}$ is zero
for $m_{1} = m_{2}$, it has a maximum at $|{\bf k}|=\sqrt{m_\1
m_\2}$. For $ |{\bf k}| \gg\sqrt{m_\1 m_\2}$, it goes like
$|V_{{\bf k}}|^2\simeq (m_\2 -m_\1)^2/(4 |{\bf k}|^2)$.

 We now use this formalism to
derive a contribution to the value of the cosmological constant
$\Lambda$.

 The connection between the vacuum energy density
$\lan\rho_{vac}\ran$ and $\Lambda$ is provided by the usual
relation $ \lan\rho_{vac}\ran= \Lambda / 4\pi G,$ where $G$ is the
gravitational constant.

To compute $\Lambda$ we can use the (0,0) component of the energy
momentum tensor in the flat space-time $T_{00}^{Flat}$. Indeed,
one can see that the temporal component of the spinorial
derivative in the FRW metric is just the standard time derivative
\cite{Blasone:2004yh}: $ D_{0}=\partial_{0}. $
 Thus, $ {\cal T}_{00} = {\cal
T}_{00}^{Flat}.$ We then obtain
\bea\
 {\cal T}_{00}(x) = \frac{i}{2}\sum_{\sigma=e,\mu}:
 \left({\bar \nu}_{\sigma}(x)\gamma_{0}
\stackrel{\leftrightarrow}{\partial}_{0}
\nu_{\sigma}(x)\right):\;=\;
 \frac{i}{2}\sum_{j=1,2}:
 \left({\bar \nu}_{j}(x)\gamma_{0}
\stackrel{\leftrightarrow}{\partial}_{0} \nu_{j}(x)\right):\eea
where $\;:...:\;$ denotes  the customary normal ordering with
respect to the mass vacuum in the flat space-time. In terms of the
annihilation and creation operators of fields $\nu_{1}$ and
$\nu_{2}$, the energy-momentum tensor $T_{00}=\int d^{3}x {\cal
T}_{00}(x)$ is given by \bea T_{00}= \sum_{r,j}\int d^{3}{\bf k}\,
\omega_{k,j}\lf(\al_{{\bf k},j}^{r\dag} \al_{{\bf k},j}^{r}+
\beta_{{\bf -k},j}^{r\dag}\beta_{{\bf -k},j}^{r}\ri). \eea Note
that $T_{00}$ is time independent.

The expectation value of $T_{00}$ in the flavor vacuum $| 0\ran_f$
gives the contribution $\lan\rho_{vac}^{mix}\ran$ of the neutrino
mixing to the vacuum energy density:
 \bea\
 {}_f\lan 0 | T_{00}|
0\ran_f = \lan\rho_{vac}^{mix}\ran \eta_{00} ~.
 \eea

Within the QFT formalism for neutrino mixing we have
 $ {}_f\lan 0 |T_{00}| 0\ran_f={}_f\lan
0(t) |T_{00}| 0(t)\ran_f $ for any t. We then obtain
\bea \non {}_f\lan 0 | T_{00}| 0\ran_f &=& \sum_{r,j}\int
d^{3}{\bf k} \, \omega_{k,j}\Big({}_f\lan 0 |\al_{{\bf
k},j}^{r\dag} \al_{{\bf k},j}^{r}| 0\ran_f + {}_f\lan 0
|\beta_{{\bf k},j}^{r\dag} \beta_{{\bf k},j}^{r}| 0\ran_f \Big),
\eea
 and
\bea\label{aspT} {}_f\lan 0 | T_{00}| 0\ran_f &=&\,8\sin^{2}\theta
\int d^{3}{\bf k}\lf(\omega_{k,1}+\omega_{k,2}\ri) |V_{\bf k}|^{2}
=\lan\rho_{vac}^{mix}\ran \eta_{00},\eea i.e. \bea\label{cc}
\lan\rho_{vac}^{mix}\ran &=& 32 \pi^{2}\sin^{2}\theta \int_{0}^{K}
dk \, k^{2}(\omega_{k,1}+\omega_{k,2}) |V_{\bf k}|^{2} , \eea
where the cut-off $\;K\;$ has been introduced. Eq.(\ref{cc}) is
our result: it shows that the cosmological constant gets a
non-zero contribution induced purely from the neutrino mixing
\cite{Blasone:2004yh}. Notice that such a contribution is indeed
zero in the no-mixing limit ($\theta = 0$ and/or $m_{1} = m_{2}$).
Moreover, the contribution is absent in the traditional
phenomenological (Pontecorvo) mixing treatment.

We may try to estimate $\lan\rho_{vac}^{mix}\ran$ by fixing the
cut-off. If we choose the cut-off proportional to the natural
scale appearing in the mixing phenomenon $\textbf{k}_{0}\simeq
\sqrt{m_{1} m_{2}}$ \cite{BV95}: using $K\sim k_{0}$, $m_{1}=7
\times 10^{-3}eV$, $m_{2}=5 \times 10^{-2}eV$,
 and
 $\sin^{2}\theta\simeq 0.3$ \cite{masses} in Eq.(\ref{cc}), we obtain
 $ \lan\rho_{vac}^{mix}\ran =0.43 \times 10^{-47}GeV^{4}$
and $ \Lambda \sim 10^{-56}cm^{-2}$ which is in agreement with the
upper bound of $\Lambda$ \cite{Zeldovic1}.
Another possible choice is to use the electro-weak scale cut-off:
$K\approx 100 GeV$. We then have $ \Lambda \sim 10^{-24}cm^{-2},$
which is, however, beyond the accepted upper bound.

In a recent paper \cite{Barenboim:2004ev}, it was suggested that,
in the context of hierarchical neutrino models, the cut-off scale
can be taken as the sum of the two neutrino masses, $K = m_{1} +
m_{2}$, resulting
 in a contribution of the right order.

In conclusion, the QFT treatment of the neutrino mixing leads to a
non zero contribution to the cosmological constant
\cite{Blasone:2004yh}.

By choosing the cut--off given by the natural scale of the
neutrino mixing phenomenon, we obtain a value of $\Lambda$ which
is consistent with its accepted upper bound. Our result discloses
a new possible, non-perturbative
 mechanism contributing to the
cosmological constant value.

Partial financial support by MURST, INFN, INFM and ESF Program
COSLAB is acknowledged.

\end{document}